\documentclass[aps,twocolumn,prb]{revtex4-2}

\usepackage{epsfig}
\usepackage{dcolumn}
\usepackage{bm}
\usepackage{amsmath}
\usepackage{tikz}
\usepackage{graphicx, graphics, color}
\usepackage{natbib}

\usepackage[colorlinks=true,citecolor=blue]{hyperref}

\ifpdf
	\DeclareGraphicsExtensions{.pdf,.png,.jpg,.mps}
\else
	\DeclareGraphicsExtensions{.eps}
\fi

\newcommand{\be}{\begin{equation}}
\newcommand{\ee}{\end{equation}}
\newcommand{\beq}{\begin{eqnarray}}
\newcommand{\eeq}{\end{eqnarray}}

\graphicspath{{./rys/}}

\begin{document}

\title{Four terminal quantum dot as an efficient rectifier of heat and charge currents}


\author{Karol I.\ Wysoki\'{n}ski}
\affiliation{Institute of Physics, M.~Curie-Sk{\l}odowska University, pl.~M.~Curie-Sk{\l}odowskiej 1, 20-031 Lublin, Poland}

\date{\today}

\begin{abstract}
 We propose an efficient method of heat rectification in a simple system consisting of a quantum
dot asymmetrically coupled to four mutually perpendicular electrodes. In such a device the Hall-like
charge and heat currents appear in response to the voltage bias or temperature difference between
one pair of electrodes. Even though both longitudinal (along the bias) and Hall-like (perpendicular
to the bias) currents are rectified under appropriate conditions, the rectification factor is typically
much bigger for the latter currents. This is true for heat and charge flow. The Hall-like currents are
predicted to exist in linear as well as non-linear transport regimes and require broken mirror symmetry but
not time reversal symmetry. The linear effect exists only in geometry which breaks two inversion
symmetries along two pairs of electrically coupled terminals. The proposed system is attainable
within current technology and provides novel platform of simultaneous heat and charge management
at the nanoscale.
\end{abstract}
\maketitle

\section{Introduction}

A rectifier is a device in which the magnitude of the
charge or heat current depends on the sign of the electric
or thermal bias. Efficient heat current rectifiers are
among the most desirable devices at the nanoscale as they
allow for heat management in miniaturised electronic devices.
Various geometries and devices have been proposed
to reach the goal. An early proposal of nanoscale
heat rectification has appeared in the context of molecular
electronics~\cite{aviram1974} and is still debated in the literature~\cite{stokbro2003,baldea2021}.

In recent years an increased interest is observed in
theoretical analysis and experimental studies of thermal
diodes~\cite{liu2019}. Besides molecules including those with
negative U centers~\cite{kiw2010}, many systems with tuned quantum
properties have been proposed as possible rectifiers.
These include inter alia recently analysed two terminal
junctions with bath particles obeying different particle
exchange statistics~\cite{palafox2022} and the use of quantum entanglement
as a possible tools to enhance rectification properties~\cite{poulsen2022}. Also various novel solid state systems and materials
have been put forward. These include heterostructures,
functionally graded or phase changing normal materials,
superconductors, etc. as reviewed recently~\cite{wong2021,malik2022}.

In the context of this paper, the rectifying devices
based on quantum dots are of special interest~\cite{dicarlo2003,inarrea2007,muller2009,kuo2010,ruokola2011,hartmann2015,rossello2017,tang2018,malz2018,lu2019,zimbovskaya2020,iorio2021,zhang2021,tesser2022}.
They consist of a single or more quantum dots coupled to
two external reservoirs. Probably the first experimental
demonstration of rectifying properties of a two terminal
quantum dot is that in Ref.~\cite{scheibner2008}  inferred from the asymmetric
line shape of the thermopower. Rectification of
both charge and heat currents in the Coulomb blockade
regime in the system with two quantum dots has
been studied in~\cite{zimbovskaya2020}. In~\cite{tesser2022} the authors concentrate on
heat rectification through quantum dots in the Coulomb
blockade regime using master equation approach. They
considered two-terminal and four-terminal devices. In
the latter case, two coupled quantum dots form a main
nanoscopic element in which each of the individual dots 
is contacted by two separate terminals.

Quantum dots play an important role in novel electronic
devices like single electron transistors~\cite{kastner1992}, heat
nano-engines~\cite{benenti2017} and many more, including building
blocks of quantum computers~\cite{loss1998,burkard2000}. Quantum dots
with large charging energy U, coupled to external metallic
leads behave like magnetic impurities in noble metals~\cite{ng1988,glazman1988}
 and at low temperature show Kondo effect~\cite{goldhabergordon1998,cronenwett1998}.

\begin{figure}[h]
\includegraphics[width=0.95\linewidth]{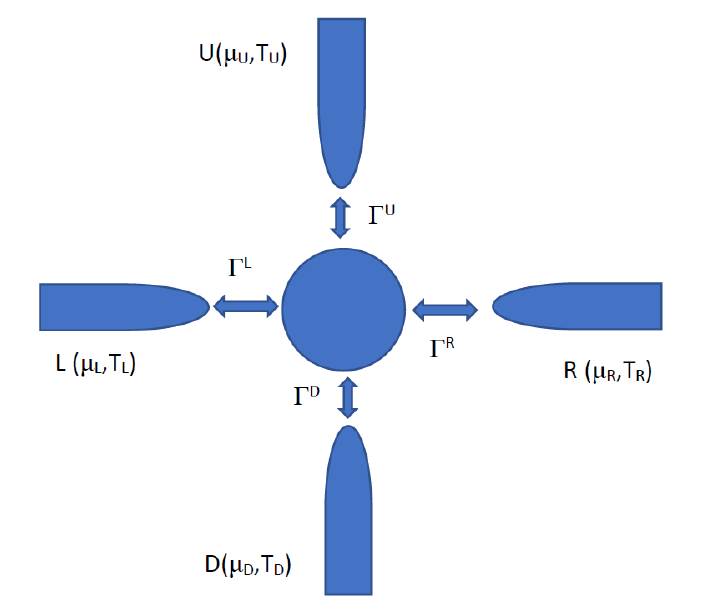}
\caption{(color online) The four terminal quantum dot geometry used in this work.
The couplings $\Gamma^\lambda$ between the central dot and the terminals  ($\lambda=L,R,U,D$) take on arbitrary values. The external leads can differ by the chemical potential $\mu_\lambda$ and/or temperature $T_\lambda$. }
\label{fig:rys1} 
\end{figure}

The device we are proposing consists of a single quantum
dot tunnel coupled to four normal electrodes in a
cross geometry as shown in Fig. (\ref{fig:rys1}). Application of the
voltage or thermal bias along $LR$ electrodes (i.e. x
direction) results in a simultaneous flow of (longitudinal)
charge and heat currents between $L$ and $R$ electrodes and
also between $U$ and $D$ electrodes (Hall-like currents) if
the system breaks mirror symmetry(-ies). A device with
non-symmetrical couplings works as an efficient rectifier.
The rectification is observed for both currents and directions.
If the interaction $U$ of electrons is non-negligible,
as it is usually the case in small structures, and if working
temperature is low enough, the device allows to study
the effect of Kondo correlations on longitudinal and Hall
- like currents.

Four terminal nano-junctions with similar geometry
have been studied previously. General symmetry properties
of phase coherent transport in the presence of magnetic
flux were derived in~\cite{buttiker1986}. The same geometry was
theoretically analysed in the context of Hall - like currents
and resistances~\cite{bulgakov1999,pichugin2000}  taking into account polarisation
of electrons and strong spin-orbit interaction in
the central region. Wei et al. studied planar four terminal
system~\cite{wei2022}  and found non-linear Hall effect induced
by the dipole of Berry curvature. The hybrid structure
consisting of a quantum dot contacted by tunnel barriers
to four electrodes including two superconducting and
two normal leads, allows~\cite{sun2000}  the control of super-current
flowing between a pair of superconducting electrodes by
the bias voltage applied to normal electrodes. The cross
geometry of the four terminal nano-junction containing an asymmetric prism -like scatterer has been studied
experimentally~\cite{song1998}. Even though our system is different
from the experimental one it features similar antisymmetric
dependence of the four terminal resistance on
the current as discussed below.

We shall analyse two possible boundary conditions:
open with floating electrodes $U$ and $D$ and open with
flow of Hall-like current. There is no Hall voltage~\cite{ashcroft}
under open boundary conditions. However, the Hall-like
currents perpendicular to the direction of the bias exist
under closed boundary conditions. These conditions
are necessary but not sufficient for the observation of
Hall-like currents. Breaking of single or two mirror symmetries
is needed. The system with broken symmetries
shows rectification properties. The rectification efficiency
may attain its maximal possible value equal unity. Tuning
the system to such hot spots allows perfect rectification
of charge or heat. We hope this property can be
verified experimentally as the fabrication of the proposed
devices is possible by the present-day technology.

The organisation of the paper is as follows. In Section (\ref{sec:system})
 we describe the studied device and its modelling.
Two possible boundary conditions are discussed
in  (\ref{sec:openbc}) and (\ref{sec:closedbc}). The results of numerical calculations
presented in Section (\ref{sec:res}) are followed by the concluding
Section (\ref{sec:summ}). Some detailed calculations of currents
flowing in the system and the on-dot Green function are
relegated to the Appendices (\ref{sec:curr-app}) and (\ref{sec:gf}).

\section{The geometry, currents and boundary conditions} \label{sec:system}
The time reversal symmetric nano-structure consisting
of a quantum dot tunnel-coupled to four normal leads
 is illustrated in Fig. (\ref{fig:rys1}). The quantum dot is understood here as the small grain with quantized spectrum, which can be modelled by a single level. 
Electrodes or terminals are macroscopic objects characterised by temperature $T\lambda = T +\Delta T_\lambda$.
They may be electrically biased ($eV_\lambda$) with chemical potential $\mu_\lambda=\mu+eV_\lambda$, where $\mu$ is an equilibrium value of the chemical potential and $T$ the equilibrium temperature
common to all electrodes. We assume $\mu = 0$.

The Hamiltonian describing the system under consideration is  a standard single-impurity Anderson model
\beq
{H}&=&\sum_{\lambda {k} \sigma}\varepsilon_{\lambda {k}}n_{\lambda {k} \sigma} + 
\sum _{\sigma} \varepsilon _{d \sigma}n_{\sigma} +U_dn_\uparrow n_\downarrow \nonumber \\
&+&\sum_{\lambda {k} \sigma} \left({V}_{\lambda {k}\sigma} c^{\dagger}_{\lambda k\sigma} d_{\sigma} 
+ {V}_{\lambda {k}\sigma}^* d^{\dagger}_{\sigma} c_{\lambda {k} \sigma}\right),
\label{eq:ham1}
\eeq
where $n_{\lambda {k} \sigma}=c^{\dagger}_{\lambda {k} \sigma}c_{\lambda {k} \sigma}$~ and $n_{\sigma}=d^{\dagger}_{\sigma} d_{\sigma}$ denote particle number operators for the leads and the dot, respectively.  The operators $c^{\dagger}_{\lambda k\sigma} (d^{\dagger}_{\sigma})$ create electrons in respective states $\lambda {k}\sigma$ $(\sigma)$ in the lead $\lambda$ (on the dot). The energies of the leads are measured from their chemical potentials $\mu_\lambda$, $\varepsilon_{\lambda k}=\varepsilon_{0\lambda k}-\mu_\lambda$, with the dependence 
of $\varepsilon_{0\lambda k}$ on $\lambda$ allowing for a different spectrum in each of the leads. 
$U_d$ denotes the energy cost of placing two electrons on a quantum dot.
In this work we neglect spin dependence of the on dot energy $\varepsilon_{d\sigma}=\varepsilon_d$
 and electron hoping amplitudes $V_{\alpha k \sigma}=V_{\alpha k}$. These approximations are relaxed in hybrid system with normal and ferromagnetic electrodes or in the presence of magnetic field.
To study the transport properties we employ the non-equilibrium Greens function
technique (for details see Appendix (\ref{sec:curr-app}) and (\ref{sec:gf})).

\subsection{Absence of Hall voltage in systems with open boundary conditions}\label{sec:openbc}

We start by assuming the bias $B$ (this denotes either voltage
$V$ or temperature $\Delta T$) between $L$ and $R$ electrodes.
For open boundary conditions chemical potentials $\mu_U$ and $\mu_D$ are obtained by requiring that the currents in those electrodes vanish: $I_U=0$ and $I_D=0$. Using equations (\ref{charge-curr-la}) and (\ref{dos}) to calculate the currents and neglecting spin dependence it is easy to show  that under arbitrary voltage bias $V_{LR}$ the vanishing of charge currents in the $U$ and $D$ electrodes requires
\be
F_D-F_U=0=\int\frac{dE}{2\pi}(f_D(E)-f_U(E))N(E).
\ee
For equal temperatures ($T_D=T_U$) the only solution is $\mu_D=\mu_U$ and no voltage appears along ($UD$)-direction, regardless the parameters of the model. This is valid for 
arbitrary voltages and temperatures, $i.e.$ in linear and
non-linear regimes. Thus under open boundary conditions
there is no 'Hall voltage' in the system and thus no
true Hall effect. This, however, does not preclude flow of
the Hall currents.
 
\subsection{The Hall-like current in the four terminal QD
with closed boundary conditions}\label{sec:closedbc}

The situation is completely different if one applies closed boundary conditions and allows for a current flow along y-direction. We bias a system along x-direction with $V=V_R-V_L$ or $\Delta T=T_R-T_L$ and calculate the current flow along the $UD$-direction. 
It has to be noted that voltage bias induces both, heat
$J^Q$ and charge $I$ currents along $LR$ and $UD$ and the same is
true for temperature bias. The currents are calculated as
$I^{LR} = (I^L - I^R)/2$ and $I^{UD} = (I^U - I^D)/2$ for charge 
and $J^{LR}_Q = (J^L_Q - J^R_Q)/2$, $J^{UD}_Q = (J^U_Q - J^D_Q )/2$ for a heat flow.

The necessary condition for the existence of currents
along unbiased $UD$ direction is broken mirror symmetry
between U and D electrodes. This is realised by
assuming different couplings to the dot. For symmetric system with $\Gamma^U=\Gamma^D$ the Hall - like currents vanish, $I^{UD}=0$ ($J^{UD}_Q=0$).
This is easily seen assuming
isothermal conditions ($T_\lambda = T$ for all $\lambda$) and linear response
regime (small voltage $V = V_L-V_R$). For symmetric distribution
of voltages $V_{L/R} = \pm V/2$ at the corresponding
terminals, one expands the Fermi functions
\be
f_\lambda(E)=f_0(E) +f_0^\prime(E)(-eV_\lambda),
\ee 
and using Eq. (\ref{charge-curr-la}) gets the Hall-like charge current
\be
I^{UD}=\frac{4e^2}{h}\frac{(\Gamma^U-\Gamma^D)[\Gamma^L-\Gamma^R]}{\sum_\lambda \Gamma^\lambda} F_0^\prime V.
\label{curr-lin}
\ee
where $F_0^\prime=\int\frac{dE}{2\pi}(-\frac{\partial f_0(E)}{\partial E})N(E)$.
Both currents vanish
for $\Gamma^L = \Gamma^R$ and/or $\Gamma^U = \Gamma^D$. This condition is relaxed
in the non-linear regime and breaking the mirror symmetry
along $UD$ is enough to get the Hall-like currents.

\section{The results}\label{sec:res}
The asymmetry of couplings plays an important role
and in most cases we shall characterise it by a single parameter
$\alpha$, which defines anisotropy of our system. It
may take arbitrary positive value but we shall study a
few representative values only. In most studied cases we
assume simple asymmetry with $\Gamma^R = \Gamma^D = \alpha\Gamma_0$ and
$\Gamma^L = \Gamma^U = \Gamma_0$, where $\Gamma_0$ is our energy unit. We also assume Boltzmann constant, Planck constant and the electron
charge as $k_B =\hbar = e = 1$. Thus energy E and other
parameters like $T$, $V$ and$\varepsilon_d$ are all measured in units of
$\Gamma_0$.

\begin{figure}
\includegraphics[width=0.85\linewidth]{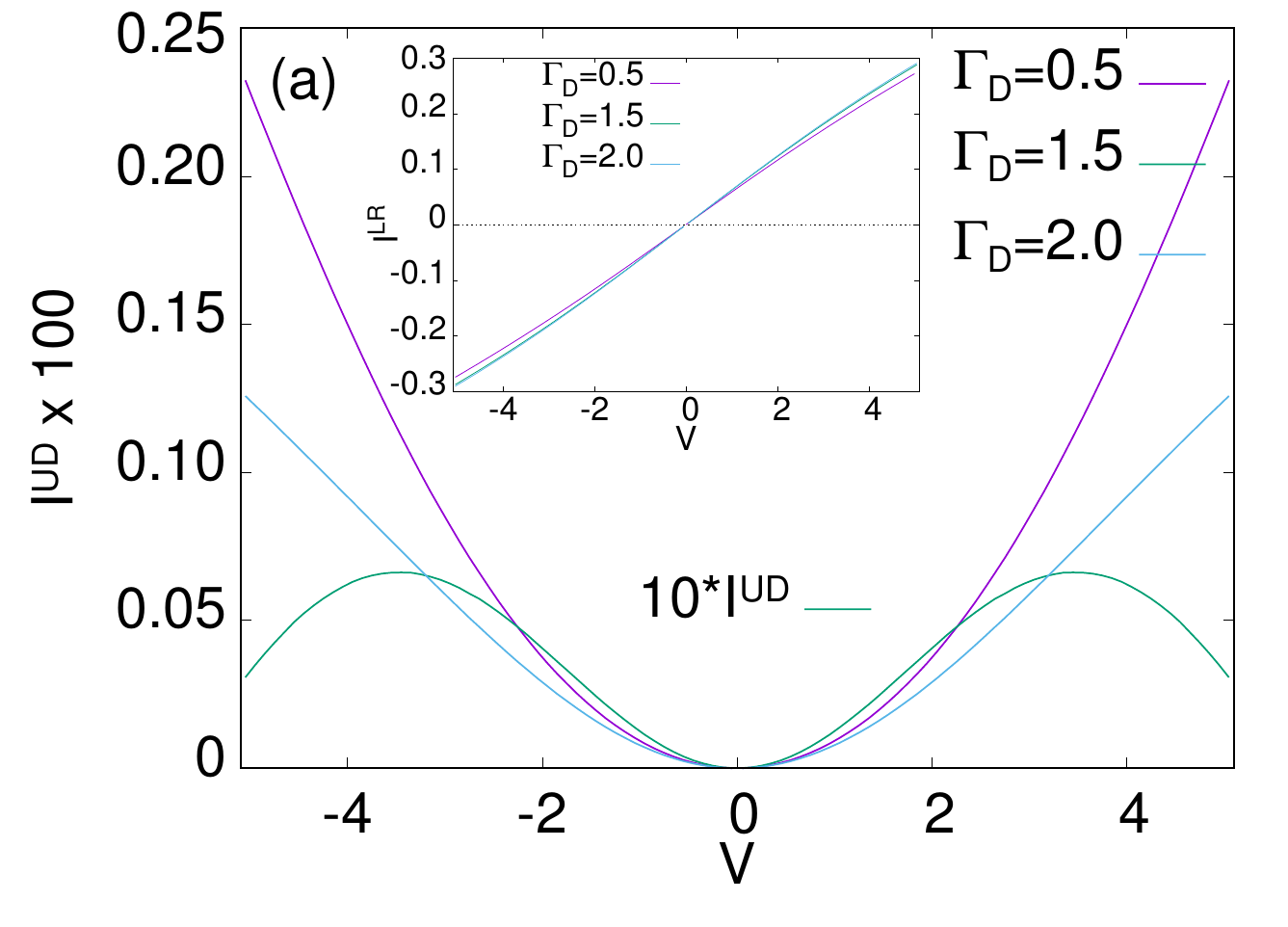}
\includegraphics[width=0.85\linewidth]{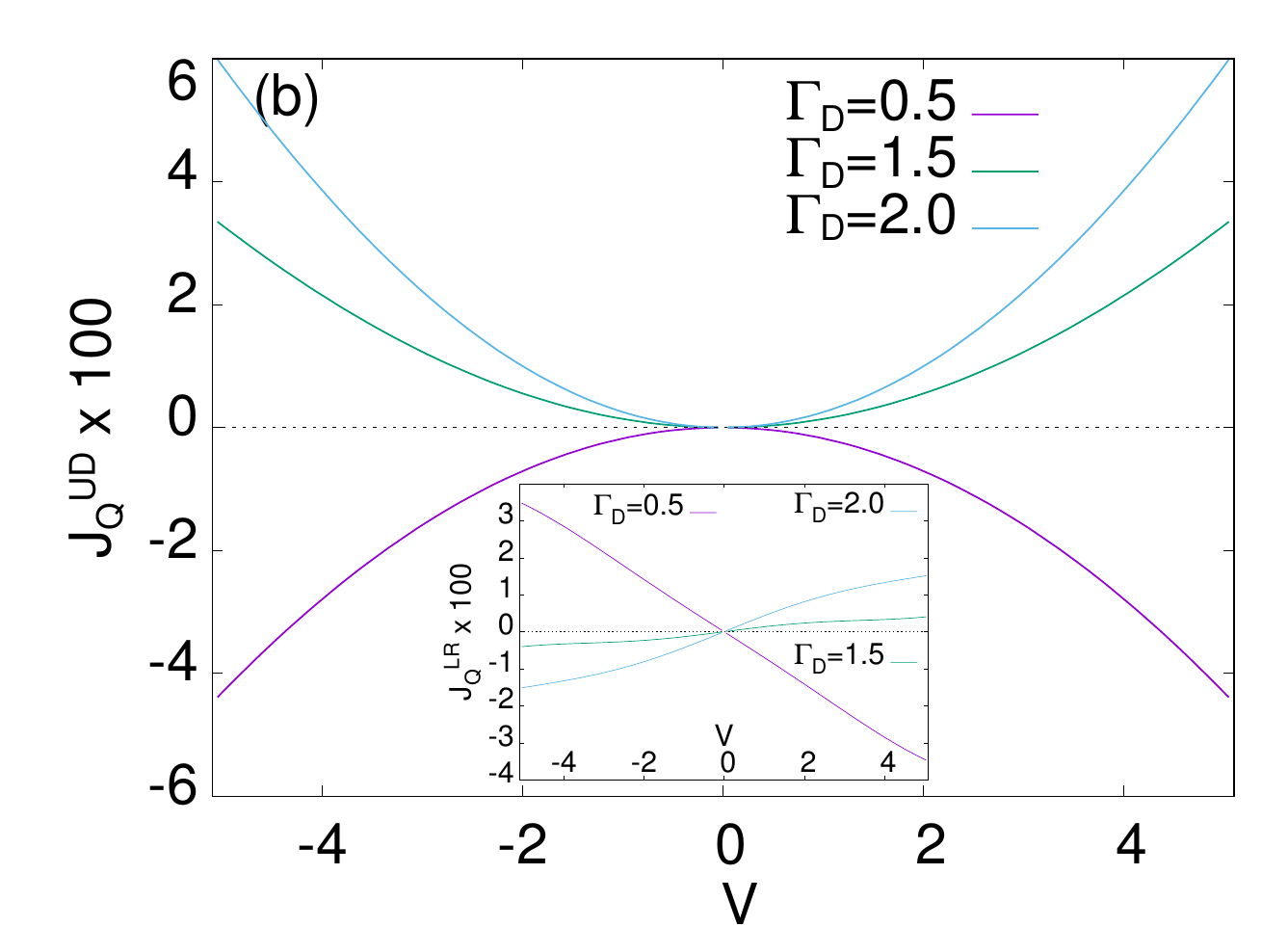}
\caption{(color online) Charge current $I^{UD}$ induced by voltage
between $LR$ terminals is a symmetric function of $V$ as visible
in panel (a). The same is true for heat current which is also
symmetric function of voltage (panel b). Insets show the corresponding
currents along $LR$ direction. These currents are
linear functions at small voltages with non-linear behaviour
observed at high bias. Other parameters $T=1$, $\varepsilon_d=-4$ and $U=12$.
}
\label{fig:rys2}
\end{figure} 

\subsection{The currents}\label{sec:curr}

Contrary to the linear regime in which existence of
Hall-like currents require breaking of two mirror symmetries,
in the non-linear regime both charge $I^{ij}$ and heat $J^{ij}_Q$ currents flow between $ij = LR$ as well as $ij = UD$ electrodes
if only a single mirror symmetry is broken along
$UD$ direction. This is illustrated in figure (\ref{fig:rys2}) for voltage
bias and three values $\Gamma^D = 0.5, 1.5, 2$ with all other
couplings equal to $\Gamma_0 = 1$. In accord with Eq. (\ref{curr-lin}) for $\Gamma^L=\Gamma^R$ the linear contributions vanish and   
the Hall currents are (at least) quadratic functions of voltage for
small $V$ . Simultaneously the longitudinal currents (both
$I^{LR}$ and $J^{LR}_Q$ ) are linear functions of $V$ for $V \rightarrow 0$ with
well visible departures from linearity at larger voltages 
(see insets in Fig. (\ref{fig:rys2}) a and b). This agrees with general
analytical results of Section (\ref{sec:closedbc}). Thermally induced
currents (not shown) exhibit the same behaviour, namely
the currents perpendicular to the bias are quadratic functions
of $\Delta T$ for small $\Delta T$. These currents do not appear
in the linear order if the system breaks the single mirror
symmetry only. In full analogy to the voltage bias the
longitudinal thermally induced currents are linear functions
of $\Delta T$ for $\Delta T \rightarrow 0$, with departures from linearity
at elevated $\Delta T$ values.

For the particular set of parameters the $UD$ heat current
is positive for $\Gamma^D > 1$ and negative for $\Gamma^D < 1$. This
is true independently of the bias as visible from panel (b)
in Fig. (\ref{fig:rys2}). Similar symmetry is valid for heat currents
along $LR$, which for $\Gamma^D > 1$ are of opposite sign to the
currents corresponding to $\Gamma^D < 1$. The flow towards $U$
is either blocked or facilitated.

\begin{figure}
\includegraphics[width=0.85\linewidth]{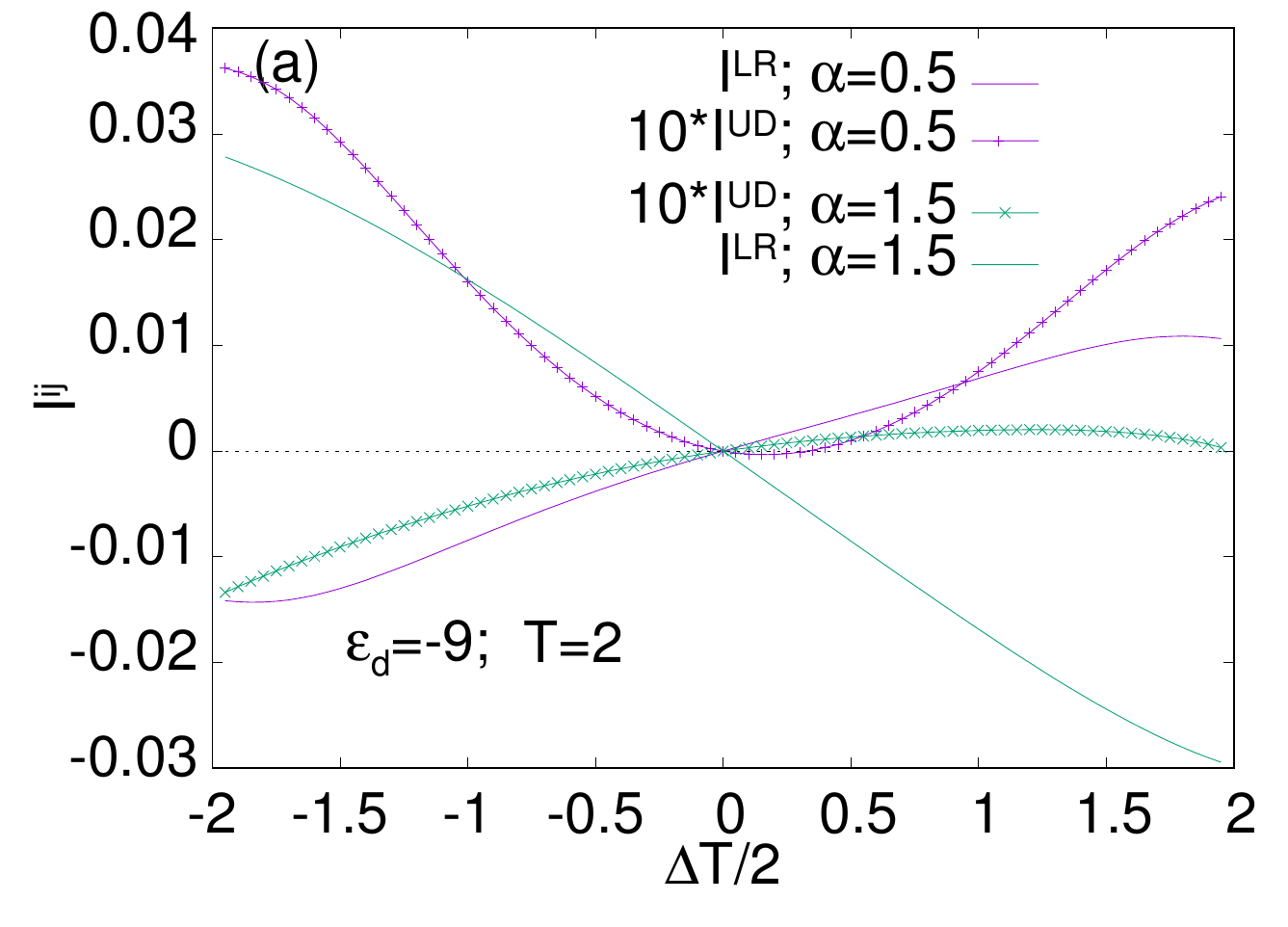}
\includegraphics[width=0.85\linewidth]{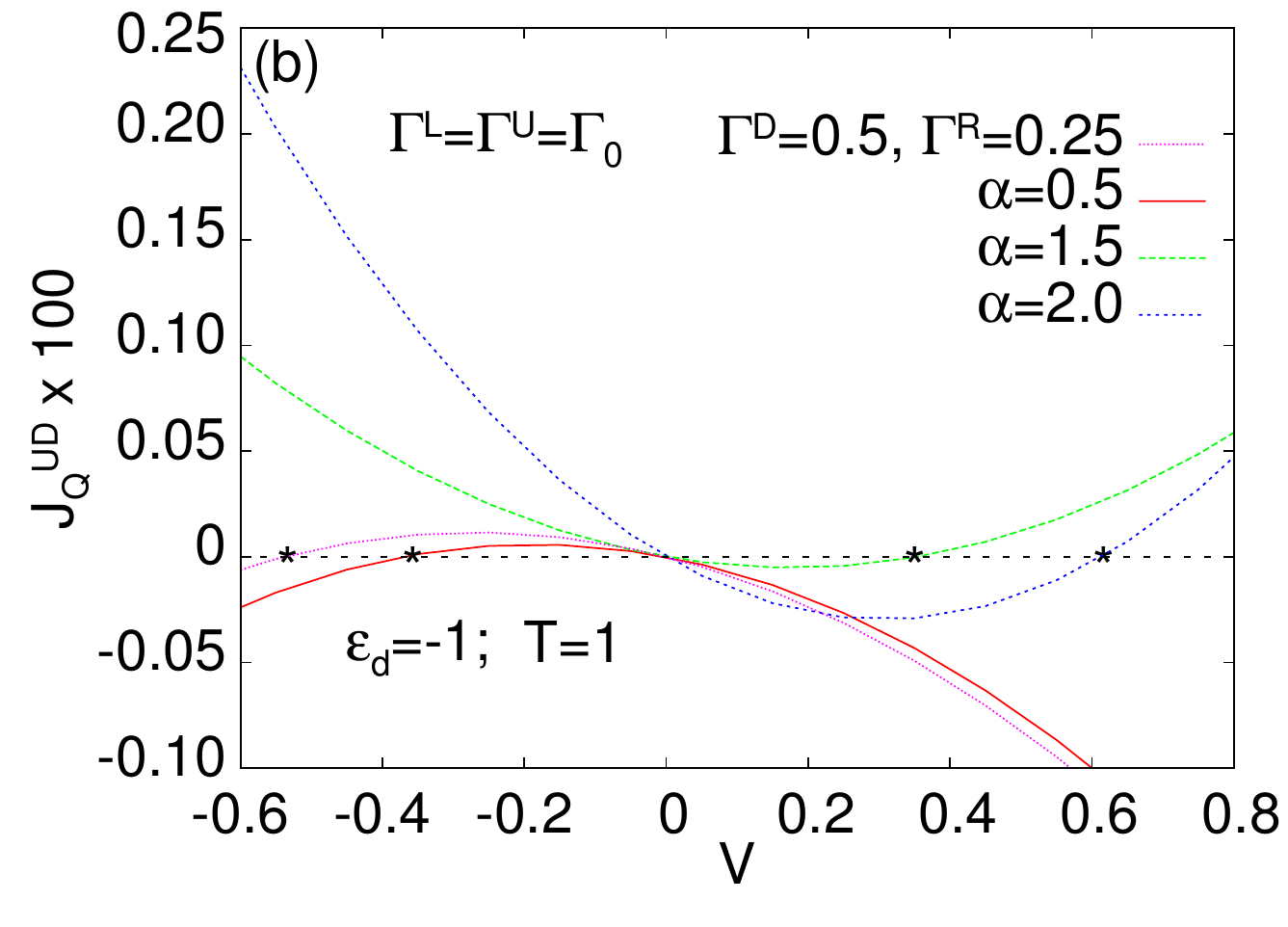}
\caption{(color online) In panel (a) we show thermally induced
longitudinal charge currents $I^{LR}$ (solid lines) and Hall-like
currents $I^{UD}$ (lines with symbols) for two values of asymmetry
$\alpha = 0.5$ and $\alpha = 1.5$ as functions of the temperature
difference. In the panel (b) the voltage dependent Hall - like
heat currents $J^{UD}_Q$  are shown together with points marked
with * where the currents vanish for one sign of voltages and
take on non-zero values for opposite sign. In both panels
$U = 12$.
}
\label{fig:rys3}
\end{figure}

The response of the system to applied bias quite generally
depends on its symmetry and the set of parameters.
To illustrate this in Fig. (\ref{fig:rys3}) we show thermally induced
charge currents in panel (a) and heat currents induced
by the voltage in panel (b). Panel (a) shows the currents
in system at temperature $T = 2$ with the quantum
dot tuned to $\varepsilon_d = -9$, and for two values of anisotropy
$\alpha = 0.5, 1.5$ and $U = 12$. The currents flowing along the
$UD$ are about an order of magnitude smaller then those
along $LR$ direction. However, one notices that the charge
Hall - like current for $\alpha = 0.5$ vanishes for $\Delta T/2 \approx 0.4$ 
but is finite for $\Delta T/2 \approx -0.4$. This is a bias for which
the rectification factor is maximal (=1) as we shall see
in the following section. In panel (b) we show Hall - like
heat currents vs. the voltage. Important fact to note is
the non-monotonous dependence of currents on the bias
and the points were they vanish. These special points
are marked by asterisks.

\subsection{Rectification}\label{sec:rect}
If mirror symmetries along $LR$ and $UD$ directions are
broken both charge and heat currents in two pairs of electrodes
depend on the sign of bias $B = V$ or $B = \Delta T$, i.e.
we find $I^{LR}(B) \ne I^{LR}(-B)$ and $I^{UD}(B) \ne I^{UD}(-B)$.
The same is true for heat currents $J^{LR}_Q(B)$ and $J^{UD}_Q(B)$.
To quantify the rectification efficiency one introduces
special parameter called rectification coefficient. One
possible definition is
\be
R^{ij} = \frac{|(|I^{ij}(B)|-|I^{ij}(-B)|)|}{[|I^{ij}(B)|+|I^{ij}(-B)|]} 
\label{rect}
\ee
With this definition the rectification factor ranges from
0 to 1. The former means no rectification, while the latter
perfect rectification. Similar rectification coefficients are
defined for heat currents $J^Q(B)$ and we denote them by
$R^{ij}_Q$ in the following.

Perfect rectification is expected in cases when the current
vanishes for one sign of the bias and attains finite
value for the other. The goal is to tune the parameters of
the system to such 'hot spots'. In the panel (b) of Fig. (\ref{fig:rys3})
such hot spots have been marked by asterisks. The rectification
coefficient takes on the maximal possible value at
those points. In general it is difficult to tune the system
and find vanishing of charge currents. However, panel
(a) of this figure shows points in which Hall-like charge
currents vanish for specific values of $\Delta T$. For $\alpha = 0.5$
the $I^{UD}$ current vanishes for $\Delta T/2$ close to 0.4, while for
$\alpha = 1.5$ the Hall - like charge current nearly vanishes for
$\Delta T/2$ close to 2. In both cases the currents are relatively
large for opposite sign of temperature bias.

\begin{figure}
\includegraphics[width=0.85\linewidth]{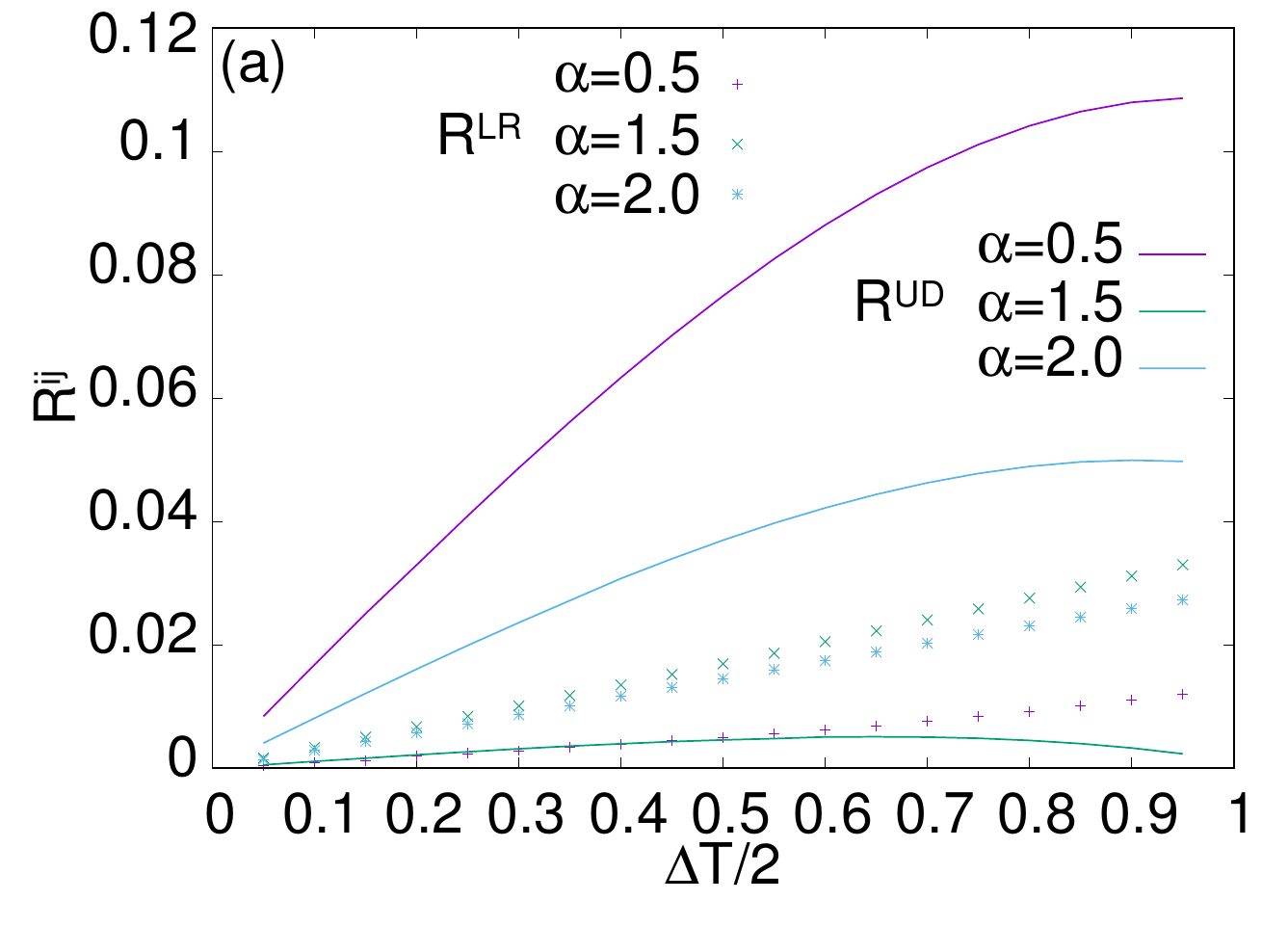}
\includegraphics[width=0.85\linewidth]{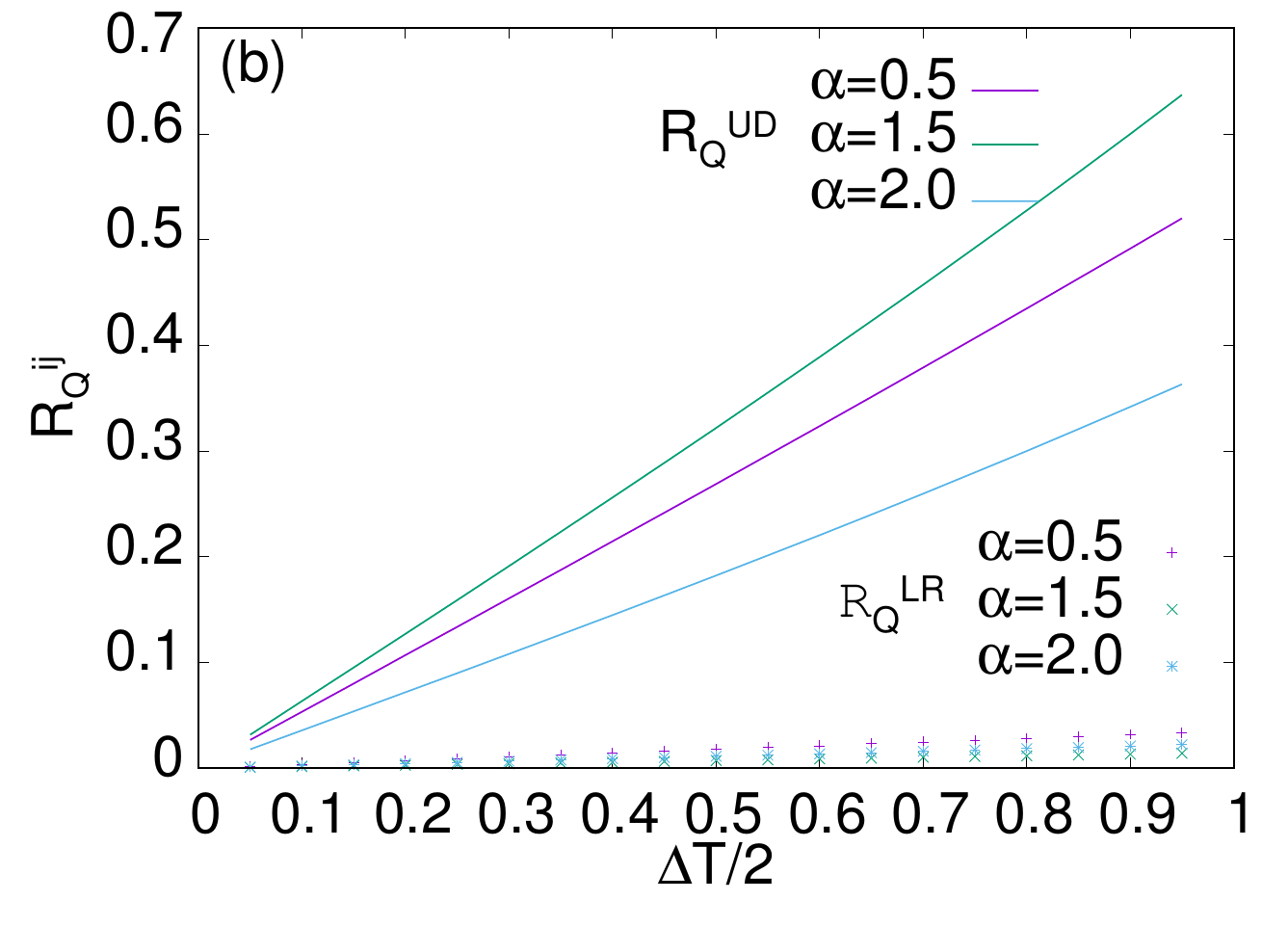}
\caption{(color online) Rectification ratios $R^{ij}$ of charge (a)
and  heat $R^{ij}_Q$ (b) currents as a function of $\Delta T/2$ for a few values of $\alpha$, voltage $V = 0$,
temperature $T = 1$, $\varepsilon_d = -4$ and $U = 12$.
}
\label{fig:rys4}
\end{figure}

The rectification coefficients for thermally induced
charge and heat currents flowing along $LR$ and $UD$
directions are plotted in Fig. (\ref{fig:rys4}). Panel (a) shows $R^{ij}$
for charge currents, while panel (b) $R^{ij}_Q$ for heat currents.
The rectification of both currents flowing along the bias
direction ($I^{LR}$ and $J^{LR}_Q$) is relatively small of order of
a few percent. For the parameters of the model presented
in the figure, the coefficient $R^{UD}$ is of the same
order or slightly higher then $R^{LR}$. Its maximal value
is about 10\%. However, the heat rectification coefficient
$R^{UD}_Q >> R^{LR}_Q$ and reaches values higher then 50\%. It 
monotonously increases with $\Delta T$.

\begin{figure}
\includegraphics[width=0.85\linewidth]{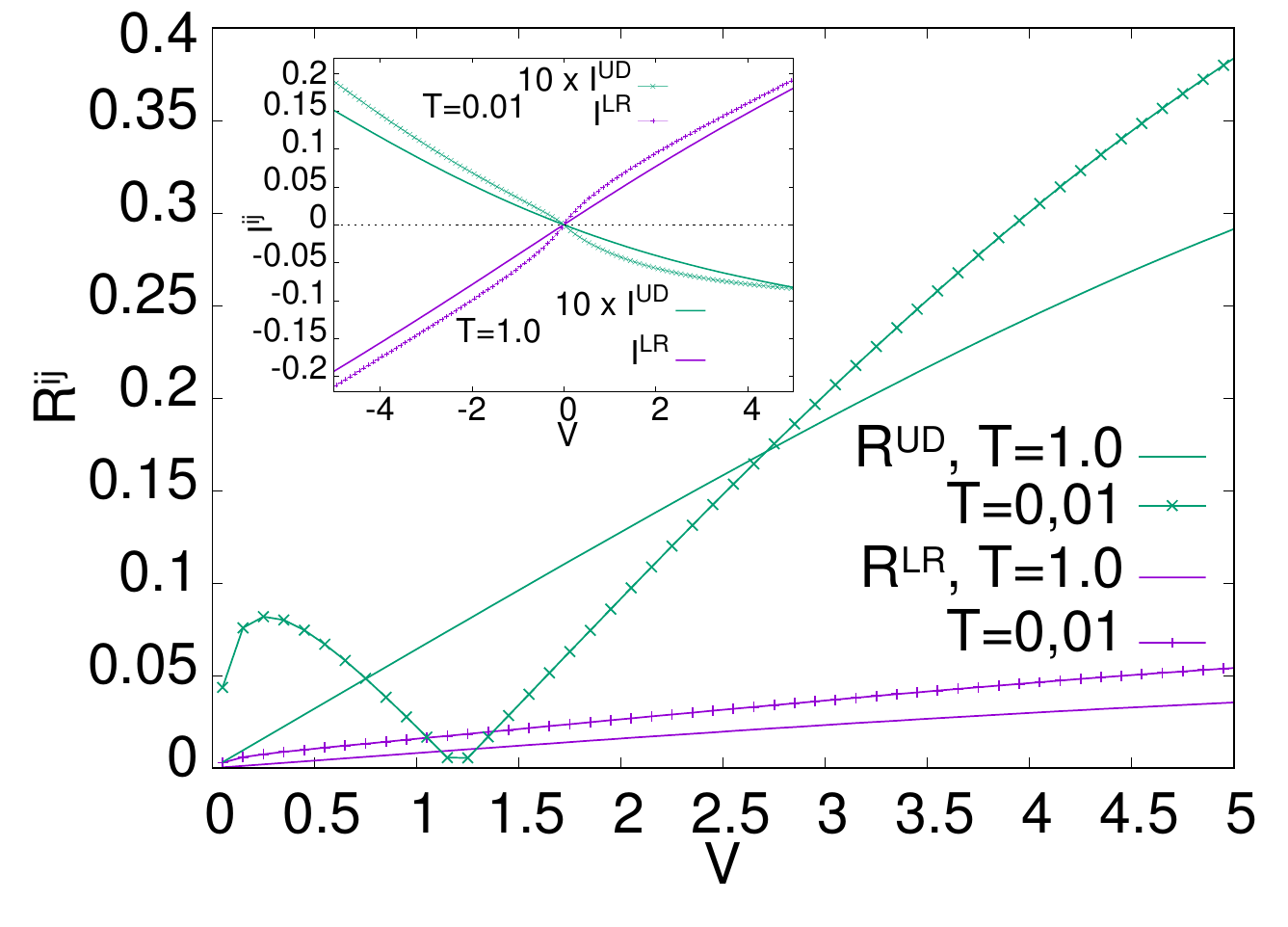}
\caption{(color online) Rectification ratios $R^{ij}$ (main panel)
and currents $I^{ij}$ (inset) as a function of voltage for two temperatures
$T = 1$ (solid lines) and $T = 0.01$ (lines with symbols).
Other parameters $\varepsilon_d = -4$ and $U = 12$.
}
\label{fig:rys5}
\end{figure}

Temperature plays an important role in our interacting
system, as the Kondo effect sets in at low temperature.
To see this we show in Fig. (\ref{fig:rys5}) results obtained
for the same system as in Fig. (\ref{fig:rys4}) for two vastly different
temperatures. We plot rectification factors $R^{ij}$ (main
panel) and the charge currents $I^{ij}$ (inset) vs. voltage $V$
for two values of temperature. For lower temperature
$T = 0.01$ the Kondo resonances are expected to appear
in the density of states. The signature of the Kondo effect
is visible in the inset as a strongly increased conductance
(the curves with symbols correspond to low
temperature) at zero voltage. Strong non-linearities at
elevated voltages combined with well visible asymmetry
of $I^{UD}$ current (at low $T = 0.01$) for voltage around
$|V^*| \approx 1.2$ result in $|I^{UD}(V^*)| = |I^{UD}(-V^*)|$ and vanishing
of $R^{UD}(V^*)$ as well as its non-monotonous dependence
on voltage. At high temperature $T = 1$ there is no
Kondo effect and one obtains continuous increase of rectification
with $V$. The rectification of longitudinal current
$R^{LR}$ is affected by the Kondo effect only quantitatively.
For the particular set of parameters its low temperature
value is roughly doubled with respect
to higher temperature but remains low, at the level of a
few percent.

The general observation is that typically the rectification
factors for longitudinal currents are small of the
order of a few percent. This agrees with previous systematic
calculations of the rectification factor in~\cite{tesser2022}  in a
two terminal single level quantum dot. The rectification
factors were small of order of a few percent like those for
longitudinal currents in the present work. In our four
terminal geometry the rectification factor for the Hall-like
currents is usually much higher and may be as large
100\%.

In view of the observation~\cite{tesser2022}  that the rectification
coefficient for a two level quantum dot attains large value
in the two terminal system it would be of interest to
extend our four terminal model to two level quantum
dot and to study how the longitudinal and Hall - like
currents and their rectification factors are affected. This
will be studied in a future work.

\subsection{Four terminal resistance in the non-linear
regime} \label{sec:resist}
For many terminal nano-junctions with currents $I^{ij}$
and voltages $V^{kl}$ measured between various pairs of electrodes
one defines~\cite{buttiker1986}  the four terminal resistances
$R_{ij,kl} =\frac{V^{kl}}{I^{ij}}$. The voltages are applied between terminals
$k$ and $l$ and currents measured between terminals $i$ and $j$.
Here we shall calculate resistances for a special couplings
which break a mirror symmetry along $y$ direction only. If
we assume couplings to $L,R$ and $U$ terminals equal to $\Gamma_0$
and only the coupling to $\Gamma^D = (0.5, 1.5, 2)\Gamma_0$ different from
others, the resulting symmetry of the model resembles
that of the nano-junction earlier studied experimentally~\cite{song1998}. 
These authors considered a four terminal structure
with asymmetric triangular prism-like scatterer placed in
its centre. The scatterer effectively blocked the flow of
charge from/to one of the terminals. In our case it is the
coupling $\Gamma^D$ which effectively blocks (if $< 1$) the current
from/to this terminal.

\begin{figure} 
\includegraphics[width=0.85\linewidth]{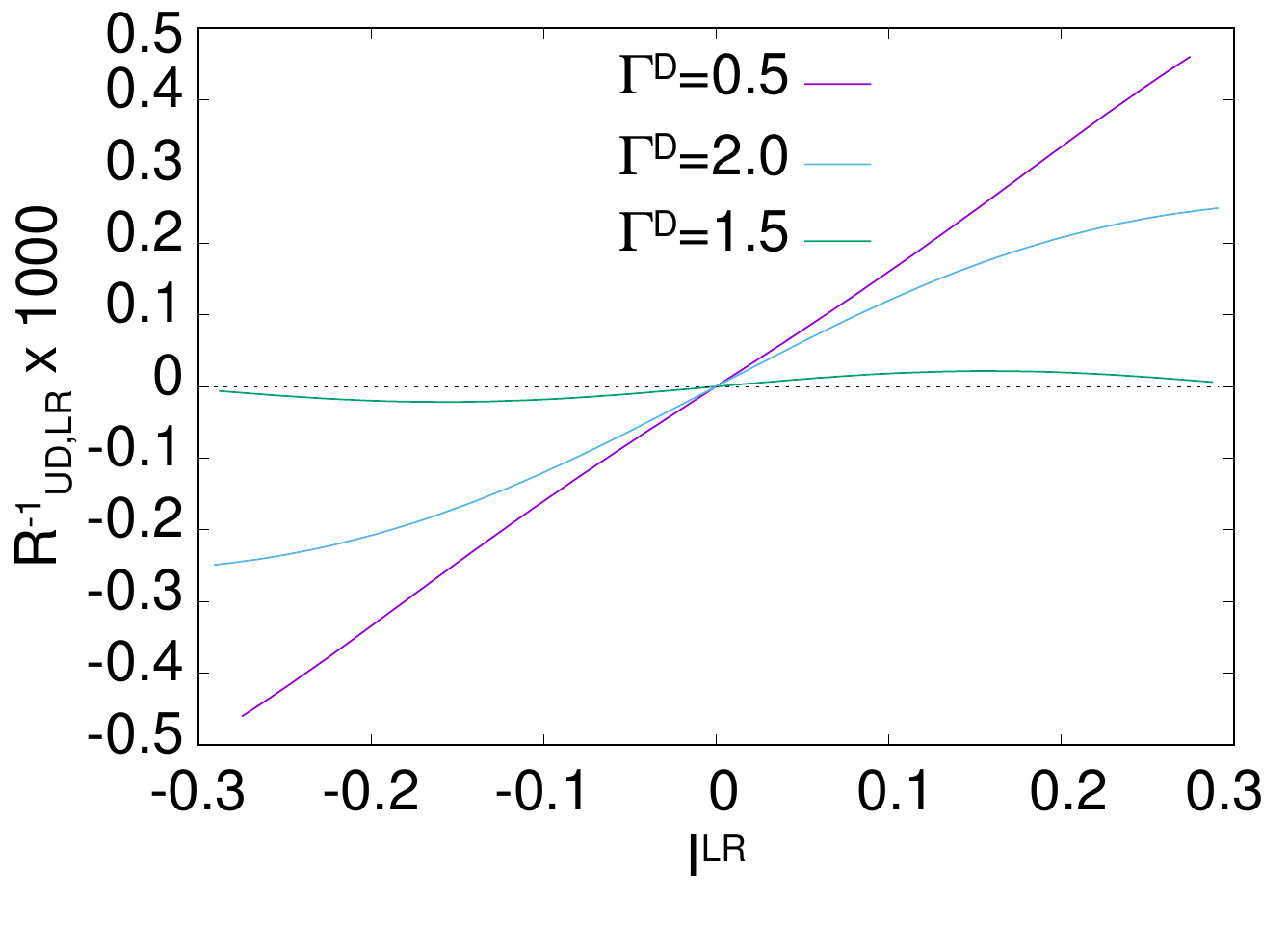}
\caption{(color online) Broken mirror symmetry along vertical
direction only $\Gamma^R = \Gamma^L = \Gamma^U = 1$ and $\Gamma^D = 0.5, 1.5, 2$ results
in the four terminal resistivity antisymmetric with respect
to the current flowing along horizontal direction. This resembles
experimental result~\cite{song1998} found in the non-linear transport
regime of nano-junction with artificial asymmetric scatterer.
}
\label{fig:rys6}
\end{figure}

It should be recalled that in the linear transport regime
the resistances $R_{ij,kl}$ do not depend on the current. In the
non-linear regime the Hall-like current $I^{UD}$ depends on
the longitudinal current $I^{LR}$ and so does  the resistance
$R_{UD,LR} = \frac{V^{LR}}{I^{UD}} = R_{UD,LR}(I^{LR})$. Due to the vanishing
of the Hall-like currents at some voltages in our set-up,
the four terminal resistance $R_{UD,LR}$ is not well
defined at those points. That is why we show in figure~(\ref{fig:rys6})
the inverse resistance $R^{-1}_{UD,LR} = \frac{I^{UD}}{V^{LR}}$ 
as a function of the current $I = I^{LR}$. It is seen that the resistance obeys
a symmetry $R_{UD,LR}(I) = -R_{UD,LR}(-I)$. Interestingly,
similar symmetry of the four terminal resistance has been
earlier observed~\cite{song1998} in the non-linear ballistic transport
in the (already mentioned) four terminal micro-junction
with triangular asymmetric (prism-like) scatterer. The
absence of voltage $V^{UD}$ between $U$ and $D$ electrodes in
our model and its presence in experimental setup is a
main reason of the perfect symmetry $R_{UD,LR}(I) = -R_{UD,LR}(-I)$ 
 in our case and only an approximate one in the
nano-junction~\cite{song1998}.

Inverse resistances $R^{-1}_{UD,LR}$ calculated in the Kondo
regime characterised by the parameters $\varepsilon_d = -4$, $U = 12$
and $T = 0.01$ display the same perfect antisymmetric dependence
(not shown) on the longitudinal current $I^{LR}$.
Needless to say that this is true for the mirror symmetry
broken along $UD$ direction in a similar way as in
experiment~\cite{song1998}.


\subsection{Gate voltage dependence and the effect of
interactions} \label{sec:gate}
In the non-linear regime the density of states (DOS)
defined in (\ref{dos}) and entering formula (\ref{f-integral}) is known to
depend crucially on the interactions  $U$ between carriers,
voltages $V_\lambda$ and temperatures $T_\lambda$ of the leads. We limit
the studies of this section to isothermal condition when
all leads have the same temperature $T_\lambda = T$. If temperature
is low enough and the on-dot level $\varepsilon_d$ is slightly
below the chemical potential(s) the Kondo peak(s) appears
in the density of states of the interacting quantum dot.
With chemical potentials $\mu = 0$ at electrodes $U$ and $D$,
$\mu_{L/R} = \mu \pm eV/2$ at the left/right electrode one expects
three Kondo peaks pinned to the chemical potentials of
the electrodes. They are visible in Fig. (\ref{fig:rys9}) shown in
Appendix (\ref{sec:gf}). Various curves in the figure correspond
to different values of $\Gamma^R = \Gamma^D = \alpha\Gamma_0$ couplings
with $\Gamma^L = \Gamma^U = \Gamma_0$. For $\varepsilon_d$ outside the Kondo regime
DOS around $E = 0$ changes smoothly with $\alpha$ and voltage
$V$ (not shown).

\begin{figure}
\includegraphics[width=0.85\linewidth]{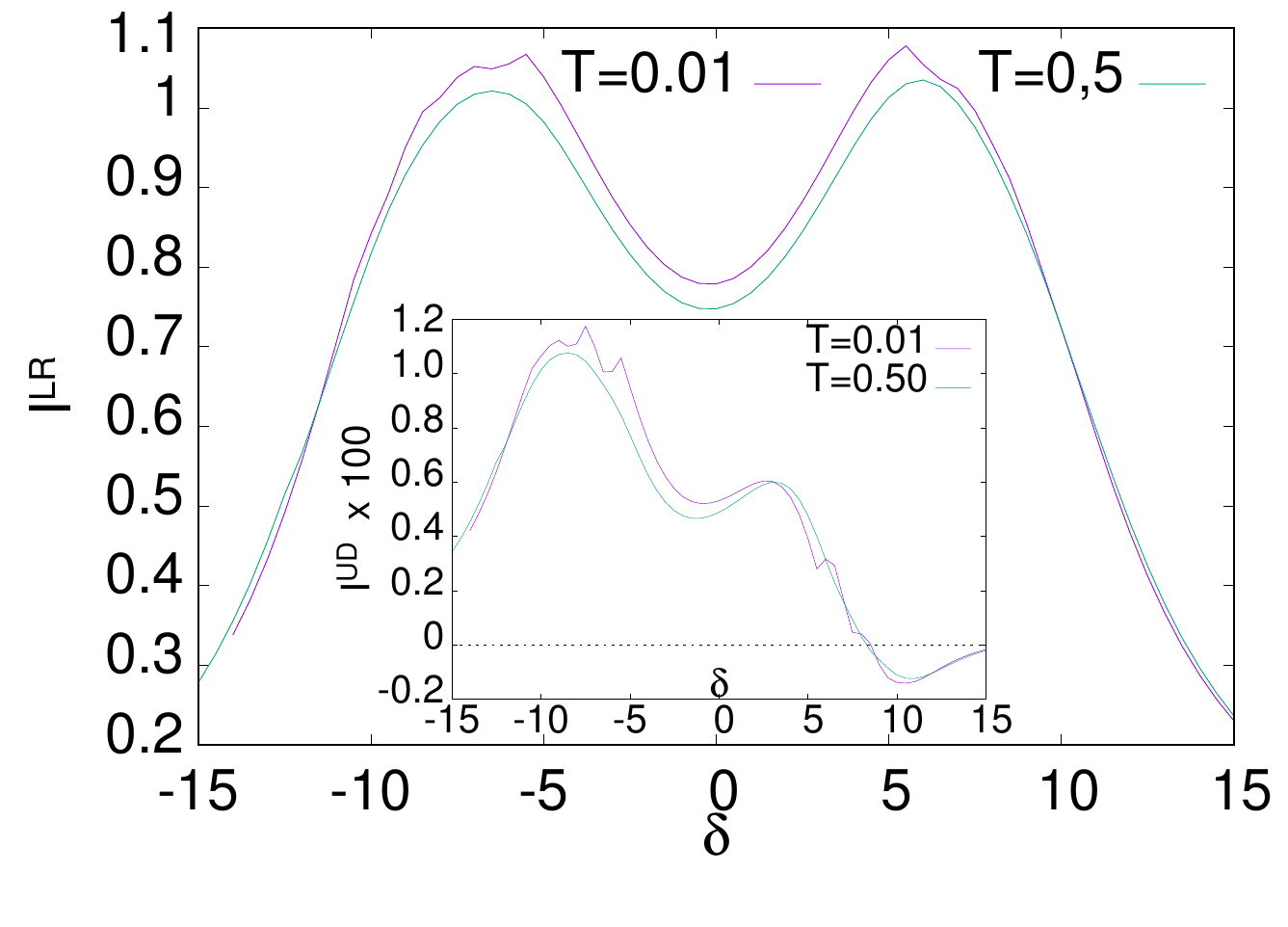}
\caption{(color online) The dependence of the longitudinal
$I^{LR}$ (main panel) and transverse $I^{UD}$ (inset) currents on the
detuning $\delta =\varepsilon_d+U/2$ (can be changed by the gate voltage) for
system with $\Gamma^L = \Gamma^U = \Gamma_0$ and $\Gamma^R = \Gamma^D = 1.5\Gamma_0$ (i.e. $\alpha = 1.5$),
$U = 16$, voltage bias $V = 4$ and for two temperatures $T =0.01, 0.5$. Signatures of the Kondo effect visible as non smooth
dependence of the currents on $\delta$ are observed for $T = 0.01$.}
\label{fig:rys7}
\end{figure}

In figure (\ref{fig:rys7}) we show the currents $I^{LR}$ (main panel)
and $I^{UD}$ (inset) as function of dot energy $\varepsilon_d$ at temperatures
below ($T=0.01$) and above ($T=0.5$) the Kondo
temperature. The dot energy can be easily changed by
gate voltage. The system breaks mirror symmetry with
respect to both $LR$ and $UD$ directions as we have assumed $\Gamma^L = \Gamma^U = \Gamma_0$ 
and $\Gamma^R = \Gamma^D = 1.5\Gamma_0$. At lower temperature one observes
signatures of the Kondo effect. These are peaks
for those gate voltages for which one expects the Kondo
resonances in the density of states (c.f. Fig. (\ref{fig:rys9})). The
effect is rather small due to the fact that we integrate the
density of states over a range ($-V/2, V/2$) around $E = 0$.
In this energy window there are three Kondo peaks including
one well pronounced and voltage independent at
$E = 0$. It is bounded with the chemical potential $\mu = 0$
of both $U$ and $D$ electrodes. Interestingly, the effect of
Kondo correlations is more pronounced in the Hall current
$I^{UD}$ as compared to longitudinal one $I^{LR}$.

\begin{figure}
\includegraphics[width=0.85\linewidth]{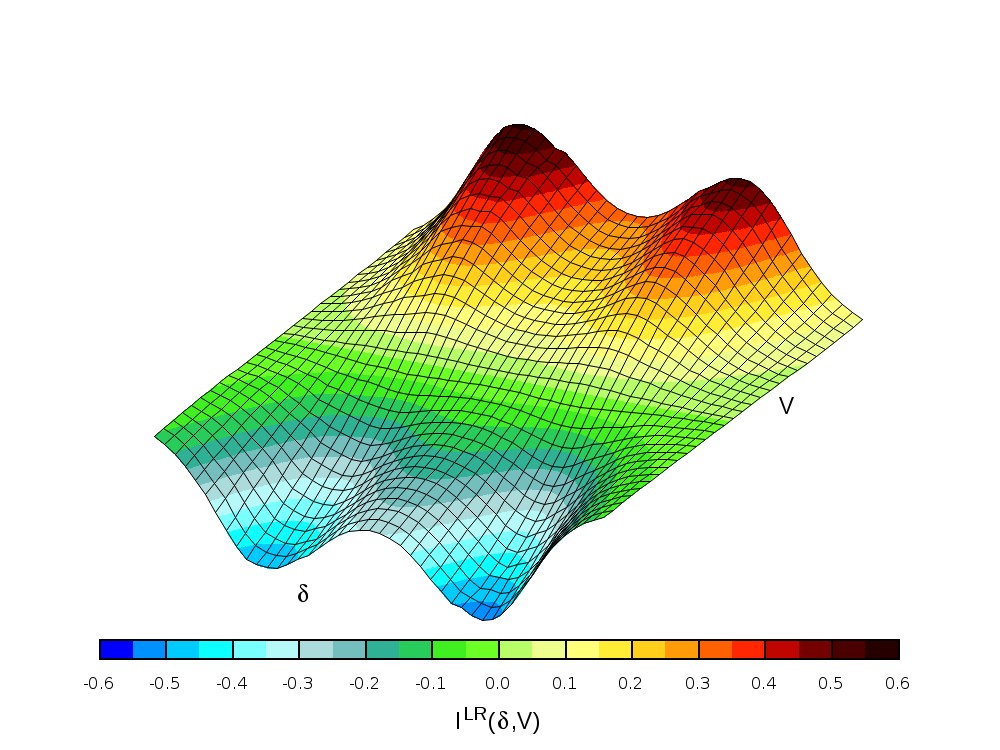}
\includegraphics[width=0.85\linewidth]{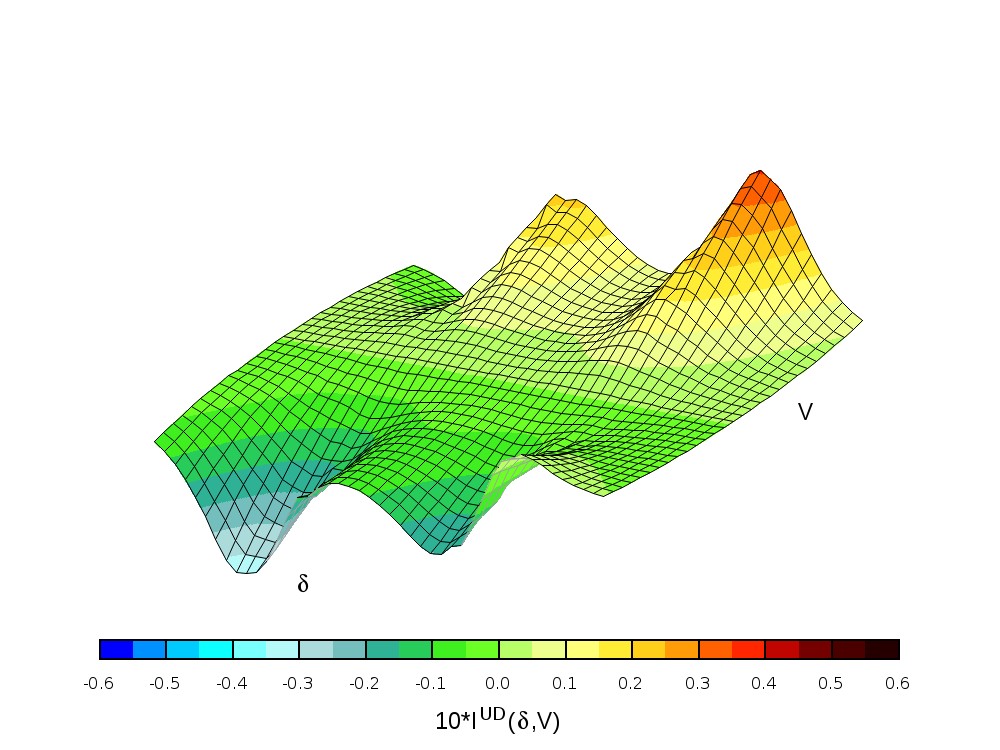}
\caption{(color online)The maps show the charge currents on
the plane ($\delta, V$ ). Upper panel illustrates longitudinal current
$I^{LR}(\varepsilon_d, V )$, while lower panel the Hall - like current
$I^{UD}(\delta, V )$. We assumed $U = 12$ and temperature $T = 0, 0.01$.
}
\label{fig:rys8}
\end{figure}

In the upper panel of colour coded Fig. (\ref{fig:rys8})  we show
the dependence of longitudinal $I^{LR}$ charge current on
$\delta = \varepsilon_d + U/2$ and voltage $V$. Lower panel shows similar
dependence for the Hall - like charge current $I^{UD}$. Both
currents are calculated for temperature in the Kondo
regime ($T = 0.01$) and for $U = 12$. Note, the same scale
for both panels, however, with Hall-like current multiplied
by the factor of 10.

\section{Summary and conclusion}\label{sec:summ}
We have studied the non-linear transport characteristics
of a system consisting of strongly interacting quantum
dot coupled to two pairs of normal leads arranged
in a cross geometry. The voltage or thermal bias applied
to one pair of the leads induces the (heat and charge)
current flow between both pairs of terminals, provided
the quantum dot is non-symmetrically coupled. Different
couplings between the leads and the dot break mirror
symmetries of the device. In the linear regime observation
of two mutually perpendicular (charge and heat) currents
in response to the appropriate bias requires breaking
of both symmetries. Beyond linear regime, breaking
the mirror symmetry along $UD$ is enough.
All currents are rectified in a system breaking both
mirror symmetries, with the rectification factors of Hall -
like currents typically much bigger then that of longitudinal
currents. Interestingly the rectification factor for heat
typically exceed that for charge. 

If only the symmetry between Hall leads (perpendicular
to those biased) is broken the resulting non-linear
four terminal resistance $R_{ij,kl} =\frac{V^{kl}}{I^{ij}}$ is an antisymmetric
function of the longitudinal current $I = I^{LR}$,
i.e. $R_{UD,LR}(I) = -R_{UD,LR}(-I)$. This is in qualitative
agreement with experimental data \cite{song1998}  on non-linear
transport in four terminal micro-junction with asymmetric
prism - like scatterer. The notable difference is that
in our system the anti-symmetry is exact while in experiment
it is approximate. It is important to recall that in
the linear regime $R_{ij,kl}$ does not depend on the current
$I = I^{ij}$.

We want also to underline that the analysed device
with a cross geometry enables the study of the Kondo effect
on both longitudinal and Hall - like currents. It turns
out that the Kondo resonance appearing in the density
of states at low temperature affects the longitudinal currents
in the four terminal system to lesser extend then in
the corresponding two terminal quantum dot geometry.
Additionally in our geometry the signatures of the Kondo
effect in the Hall -like current are more pronounced then
in the longitudinal one.

Our results demonstrate a new route to achieve the efficient rectification of longitudinal
and Hall - like (heat and charge) currents. The
proposed system is attainable within current technology
and provides a novel platform of simultaneous heat and
charge management at the nanoscale.

\acknowledgements{The work reported here has been supported by the
M. Curie-Sk\l{}odowska University and the National Science
Center, Poland (``Weave'' programme) through grant no. 2022/04/Y/ST3/00061.}

\appendix \label{append}
\section{The current in the system}\label{sec:curr-app}
Calculating the currents we follow the general definitions
relying on equation of motion of the number and
energy operators~\cite{meir1992,meir1994}. The required Green functions
are obtained in a standard way~\cite{haug-jauho1996,eckern2020,eckern2021}. Additionally
we are assuming the wide band limit when the couplings
$\Gamma^\lambda_\sigma(E) = 2\pi\sum_k |V_{\lambda k \sigma}|^2\delta (E-\varepsilon_{\lambda k})=\Gamma^\lambda_\sigma$ do not depend on energy. However, we keep here the spin dependence of adequate parameters. In this limit one finds the exact relation (for details see~\cite{eckern2021,lavagna2015})
\beq  
 \langle d^\dagger_\sigma d_\sigma\rangle&=& -i\int \frac{dE}{2\pi} G_\sigma^<(E) 
 \label{DD-noneq1} \\
&=&i\int \frac{dE}{2\pi}\frac{\sum_\lambda \Gamma_\sigma^\lambda f_\lambda(E)}{\sum_\lambda\Gamma_\sigma^\lambda}
[G^r_\sigma(E)-G^a_\sigma(E)], \nonumber
\eeq
which allows to write the charge and heat currents flowing
out of the $\lambda$ electrode as
\beq \label{charge-curr-la}
I_{\lambda}&=&\frac{2e}{\hbar} \int\frac{dE}{2\pi}\sum_\sigma \Gamma_\sigma^{\lambda}
\nonumber \\
&\times& \frac{\sum_{\lambda'} \Gamma^{\lambda'}_\sigma(f_{\lambda'}(E)-f_{\lambda}(E))}{\sum_{\lambda'} \Gamma^{\lambda'}_\sigma} 
\mathrm{Im} G_\sigma^{r}(E), \\
J_{\lambda}&=&\frac{2e}{\hbar} \int\frac{dE}{2\pi}\sum_\sigma\Gamma_\sigma^{\lambda}(E)
(E-\mu_\lambda)  \nonumber \\
&\times& \frac{\sum_{\lambda'} \Gamma^{\lambda'}_\sigma(f_{\lambda'}(E)-f_{\lambda}(E))}{\sum_{\lambda'} \Gamma^{\lambda'}_\sigma} 
\mathrm{Im} G_\sigma^{r}(E).
\label{heat-curr-la}
\eeq
These expressions can be used for calculating the currents in an arbitrary system consisting of the central dot and several terminals under appropriate boundary conditions.  Note, the Fermi-Dirac distribution function $f_{\lambda}(E))$ depends on the electrode $\lambda$ $via$ its chemical potential $\mu_\lambda=\mu+eV_\lambda$ or applied voltage $V_\lambda$ and temperature $T_\lambda$. The common value of the chemical potential and temperature of the system in equilibrium is denoted $\mu, T$. 
It is convenient to express the currents in terms of auxiliary function
\be
F_{\lambda \sigma}=\int\frac{dE}{2\pi}f_\lambda(E)N_\sigma(E),
\label{f-integral}
\ee
\be
F_{\lambda \sigma}^Q=\int\frac{dE}{2\pi}(E-\mu_\lambda)f_\lambda(E)N_\sigma(E),
\label{fQ-integral}
\ee
respectively. The auxiliary function  
\be
N_\sigma(E)=-\frac{1}{\pi}Im G^r(E, \{V_\lambda\})
\label{dos}
\ee
 denotes density of states (DOS) on the dot for spin $\sigma$ electrons. In general this quantity depends on the voltages $\{V_\lambda\}$ and temperatures $T_\lambda$ of all terminals. 
\begin{figure}
\includegraphics[width=0.85\linewidth]{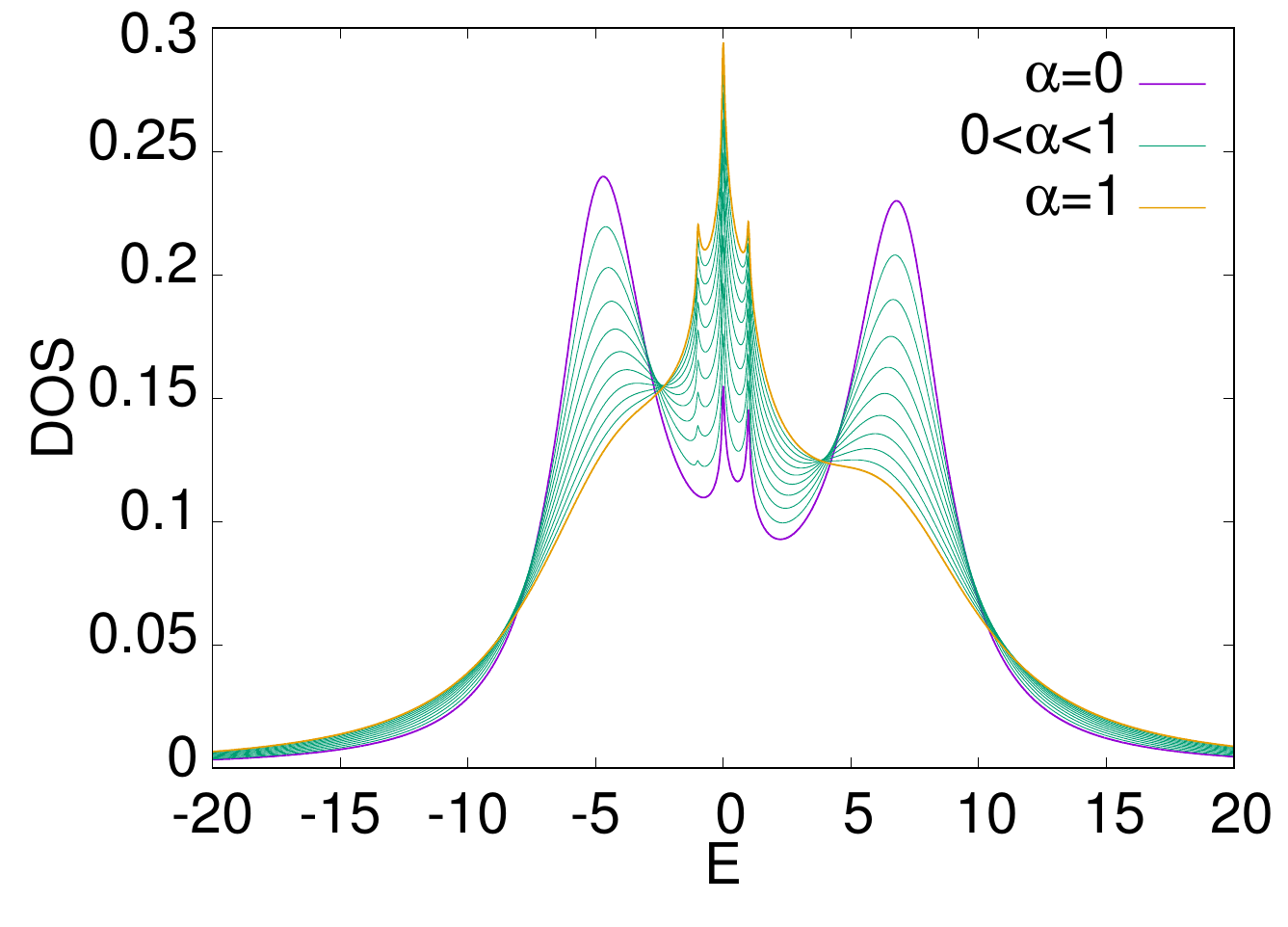}
\caption{(color online)The evolution of the on-dot density
of states plotted as a function of energy with changing the
parameter $\alpha$ describing the couplings between the dot and
right ($R$) and down ($D$) terminals with $\Gamma^R = \Gamma^D = \alpha \Gamma_0$,
$\Gamma^L = \Gamma^U = \Gamma_0$. The parameters take on the following values:
the on-dot energy $\varepsilon_d = -5.0$, the source-drain voltage $V = 2$,
temperature $T = 0.01$ and the interaction $U = 12$. 
Note that the region of width $V$ around the chemical potential $\mu =
0$ is important as it contributes to the currents in particular
terminals.}
\label{fig:rys9}
\end{figure}

\section{The Green function} \label{sec:gf} 
For completeness we recall the formulae for the on-dot Green function which qualitatively correctly describe the Kondo effect. One uses  standard equation of motion technique\cite{lavagna2015,eckern2020,eckern2021} and finds the Green function  
\begin{equation}
\label{sol-gf}
\langle\langle d_\sigma|d^\dagger_\sigma\rangle\rangle_E= 
\frac{1+I_d(E)[\langle n_{\bar{\sigma}}\rangle+b_{1\bar{\sigma}}-b_{2\bar{\sigma}} ]}
{E-\varepsilon_\sigma-\Sigma_{0\sigma}+\Sigma_t(E)}, 
\end{equation}
where
\begin{equation}
\Sigma_t(E)=I_d(E)[\Sigma_1^T+\Sigma_2^T-(b_{1\bar{\sigma}}-b_{2\bar{\sigma}})\Sigma_{0\sigma}],
\end{equation}

\begin{equation}
I_d(E)=\frac{U}{E-\varepsilon_\sigma-U-\Sigma_{0\sigma}-\Sigma_\sigma^{(1)}-\Sigma_\sigma^{(2)}}.
\label{sol-sp2}
\end{equation}

The  various pieces of the  self-energy are supplemented by the inverse life-times $i\gamma_{\bar{\sigma}_1/2}$ of the single particle $\bar{\sigma}$, respectively  two-particle $2$ state and read  
\begin{equation}
b_{1\bar{\sigma}}(E) = \int \frac{d\varepsilon}{2\pi}\frac {\sum_{\lambda} \Gamma^{\lambda}_{\bar{\sigma}} f_{\lambda} (\varepsilon) 
\langle\langle d_{\bar{\sigma}}|d^{\dagger}_{\bar{\sigma}}\rangle\rangle ^{a}_{\varepsilon}}
{E-\varepsilon-\varepsilon_1+i\tilde{\gamma}^{\bar{\sigma}}_1},
\end{equation}
\begin{equation}
b_{2\bar{\sigma}}(E) = \int \frac{d\varepsilon}{2\pi}\frac {\sum_{\lambda} \Gamma^{\lambda}_{\bar{\sigma}} f_{\lambda} (\varepsilon) 
\langle\langle d_{\bar{\sigma}}|d^{\dagger}_{\bar{\sigma}}\rangle\rangle ^{a}_{\varepsilon}}
{E+\varepsilon-\varepsilon_2+i\tilde{\gamma}_2},
\end{equation}
\begin{equation}
\Sigma^{T}_{1\sigma}(E)=\int\frac{d\varepsilon}{2\pi}\frac{\sum_{\lambda}\Gamma^{\lambda}_{\bar{\sigma}} f_{\lambda}(\varepsilon)[1+\frac{i}{2}
\Gamma_{\bar{\sigma}}\langle\langle d_{\bar{\sigma}}|d^{\dagger}_{\bar{\sigma}}\rangle\rangle^{a}_{\varepsilon}]}
{E-\varepsilon-\varepsilon_1+i\tilde{\gamma}^{\bar{\sigma}}_1},
\end{equation}
\begin{equation}
\Sigma^{T}_{2\bar{\sigma}}(E) =\int \frac{d\varepsilon}{2\pi} \frac{\sum_\lambda \Gamma^{\lambda}_{\bar{\sigma}} 
 f_{\lambda}(\varepsilon) [1-\frac{i}{2}\Gamma_{\bar{\sigma}}\langle\langle d_{\bar{\sigma}}|d^{\dagger}_{\bar{\sigma}}\rangle\rangle^{r}_{\varepsilon}]}{E +\varepsilon - \varepsilon_2+ i\tilde{\gamma}_2}.
\end{equation}
In the above we have introduced $\varepsilon_1=\tilde{\varepsilon}_\sigma-\tilde{\varepsilon}_{\bar{\sigma}}$,
and $\varepsilon_2=\tilde{\varepsilon}_\sigma+\tilde{\varepsilon}_{\bar{\sigma}}+U$. The subscripts 1 and 2 refer to the excited 1- and 2-electron states of the dot, respectively.
The symbols $a/r$ denote advanced/retarded Green function. 
The self-consistency requires that input dot occupation $\langle n_{\bar{\sigma}}\rangle$ equals that obtained from $G_{\bar{\sigma}}^r(E)$ in the consecutive iteration step with a given accuracy. As noted earlier, there exists exact relation 
\begin{equation}
 \langle n_\sigma \rangle= \int dE
\frac{\sum_\lambda\Gamma_\sigma^\lambda f_\lambda(E)}{\sum_\lambda \Gamma_\sigma^\lambda}
(-\frac{1}{\pi}) Im G_\sigma^r(E), 
\label{n-noneq} 
\end{equation} 
valid for energy independent couplings;  
$\Gamma_\sigma^{\lambda}(E) \equiv \Gamma^{\lambda}_\sigma $.
If this condition is violated, as it might be the case in graphene 
\cite{wysokinskimm2012,wysokinskimm2013,wysokinskimm2014}, 
hybrid systems with one (or both) of the electrodes being a superconductor, 
$e.g.$, d-wave \cite{polkovnikov2002} one, other approaches are needed.

The inverse lifetimes $\tilde{\gamma}_\alpha$
of the excited states $\alpha=|\sigma\rangle,|\uparrow,\downarrow\rangle$ stem from higher 
order processes~\cite{meir1994,lavagna2015}. They can be calculated up to the desired order $via$ the generalized Fermi rule as
\begin{equation}
\tilde{\gamma}_\alpha=2\pi\sum_{|f\rangle}|\langle T(E_\alpha)\rangle |^2\delta(E_\alpha-E_f),
\label{gen-fermi-rule}
\end{equation}
with $T(E)=\hat{V}+\hat{V} g(E) \hat{V} +\cdots$ the scattering matrix, where $\hat{V}$ denotes the
part of the Hamiltonian describing the coupling between quantum dot and reservoirs. 
In the discussed approach, one also replaces $\varepsilon_d$ by $\tilde{\varepsilon}_d$, to be calculated self-consistently from 
\begin{equation}
\tilde{\varepsilon}_d=\varepsilon_d+\Sigma_1^T(\tilde{\varepsilon}_d)+\Sigma_2^T(\tilde{\varepsilon}_d).
\label{edn}
\end{equation} 
Finally, the self-energies $\Sigma_\sigma^{(1,2)}$ are equal to $\Sigma_{0\sigma}$ 
for $i\tilde{\gamma}^\alpha_{1,2}=i0^+$; however, for arbitrary values of $i\tilde{\gamma}^\alpha_{1,2}$ they have to be calculated directly from 
\begin{equation}
\Sigma^{(1,2)}_\sigma(E)=\sum_{\lambda} \Gamma^{\lambda}_{\bar{\sigma}}\int \frac{d\varepsilon}{2\pi}\frac {1}
{E \mp \varepsilon-\varepsilon_{1,2}+i\tilde{\gamma}^{\bar{\sigma}}_{1,2}}.
\end{equation}
The on-dot density of states is obtained from the retarded
Green function. It is defined by Eq. (\ref{dos})  and 
shown in figure (\ref{fig:rys9}) as a function of energy for a number
of values of the asymmetry factor $\alpha$ (c.f its definition
in Section (\ref{sec:res})  affecting couplings between the dot
and leads. One observes three Abrikosov-Suhl resonances
pinned to Fermi levels of the left $\mu_L$, right $\mu_R$ and up
and down $\mu_U = \mu_D = \mu = 0$ electrodes. Terminal $R$
with $\mu_R = -1.0$ is decoupled from the dot for $\alpha = 0$ so
the corresponding density of states shows two Kondo resonances:
one at $\mu_U = 0$ and other at $\mu_L = +1.0$. With
increasing $\alpha$ the third Kondo resonance appears around
energy $E = -1 = \mu_R$. It has to be noted that the Kondo
peak at zero energy and two such structures at $V_{L/R}$ 
make the dependence of currents on gate voltage more 
complicated in comparison to the two terminal case as the
central resonance is always present and only its weight
changes with parameters.


\begin{thebibliography}{99}
\bibitem{aviram1974} Arieh Aviram, Mark A. Ratner, {\it Molecular rectifiers}, Chemical Physics Letters \textbf{29}, 277 (1974).

\bibitem{stokbro2003}  Kurt Stokbro, Jeremy Taylor, Mads Brandbyge, {\it Do
Aviram-Ratner Diodes Rectify?} J. Am. Chem. Soc. \textbf{125},
3674 (2003).

\bibitem{baldea2021}  Ioan Baldea, {\it Why Asymmetric Molecular Coupling to
Electrodes Cannot Be at Work in Real Molecular Rectifiers},
Phys. Rev. B \textbf{103}, 195408 (2021).

\bibitem{liu2019}  Hexin Liu, Haidong Wang, Xing Zhang, {\it A Brief Review
on the Recent Experimental Advances in Thermal Rectification
at the Nanoscale}, Appl. Sci. \textbf{9}, 344 (2019).

\bibitem{kiw2010}  Karol Izydor Wysoki\'nski, {\it Thermal transport of molecular
junctions in the pair tunneling regime}, Phys. Rev. B \textbf{82},
115423 (2010).

\bibitem{palafox2022} Stephania Palafox, Ricardo Rom\'an-Ancheyta, Baris
Cakmak, \"Ozg\"ur E. M\"ustecaplioglu, {\it Heat transport and
rectification via quantum statistical and coherence asymmetries},
Phys. Rev. E \textbf{106}, 054114 (2022).

\bibitem{poulsen2022}  Kasper Poulsen, Alan C. Santos, Lasse B. Kristensen,
Nikolaj T. Zinner, {\it Entanglement-enhanced quantum rectification},
Phys. Rev. A \textbf{105}, 052605 (2022).

\bibitem{wong2021}  M.Y. Wong, C.Y. Tso, T.C. Ho, H.H. Lee, {\it A review of
state of the art thermal diodes and their potential applications},
International Journal of Heat and Mass Transfer
\textbf{164}, 120607 (2021).

\bibitem{malik2022}  F K Malik and K Fobelets, {\it A review of thermal rectification
in solid-state devices}, J. Semicond., \textbf{43}, 103101
(2022).

\bibitem{dicarlo2003}  L. DiCarlo, C. M. Marcus, J. S. Harris, Jr., {\it Photocurrent, Rectification, and Magnetic Field Symmetry of Induced
Current through Quantum Dots}, Phys. Rev. Lett.
\textbf{91}, 246804 (2003).

\bibitem{inarrea2007}  Jes\'us I\~{n}arrea, Gloria Platero, Allan H. MacDonald, {\it Electronic
transport through a double quantum dot in the
spin-blockade regime: Theoretical models}, Phys. Rev. B
\textbf{76}, 085329 (2007).

\bibitem{muller2009}  C. R. M\"uller, L. Worschech, S. Lang, M. Stopa, A.
Forchel, Quantized rectification in a quantum dot nanojunction,
Phys. Rev. B \textbf{80}, 075317 (2009).

\bibitem{kuo2010}  David M.-T. Kuo, Yia-chung Chang, {\it Thermoelectric and
thermal rectification properties of quantum dot junctions},
Phys. Rev. B \textbf{81}, 205321 (2010).

\bibitem{ruokola2011}  Tomi Ruokola, Teemu Ojanen, {\it Single-electron heat diode:
Asymmetric heat transport between electronic reservoirs
through Coulomb islands}, Phys. Rev. B \textbf{83}, 241404
(2011).

\bibitem{hartmann2015}  F. Hartmann, P. Pfeffer, S. H\"ofling, M. Kamp, L.
Worschech, {\it Voltage Fluctuation to Current Converter
with Coulomb-Coupled Quantum Dots}, Phys. Rev. Lett.
\textbf{114}, 146805 (2015).

\bibitem{rossello2017} Guillem Rossell\'o, Rosa L\'opez, Rafael S\`anchez, {\it Dynamical Coulomb blockade of thermal transport}, Phys. Rev. B
\textbf{95}, 235404 (2017).

\bibitem{tang2018}  Gaomin Tang, Lei Zhang, Jian Wang, {\it Thermal rectification
in a double quantum dots system with a polaron
effect}, Phys. Rev. B \textbf{97}, 224311 (2018).

\bibitem{malz2018}  Daniel Malz, Andreas Nunnenkamp, {\it Current rectification
in a double quantum dot through fermionic reservoir engineering},
Phys. Rev. B \textbf{97}, 165308 (2018).

\bibitem{lu2019}  Jincheng Lu, Rongqian Wang, Jie Ren, Manas Kulkarni,
Jian-Hua Jiang, {\it Quantum-dot circuit-QED thermoelectric
diodes and transistors}, Phys. Rev. B \textbf{99}, 035129
(2019).

\bibitem{zimbovskaya2020}  Natalya A Zimbovskaya, {\it Charge and heat current rectification
by a double-dot system within the Coulomb blockade
regime}, J. Phys.: Condens. Matter \textbf{32}, 325302 (2020).

\bibitem{iorio2021}  A. Iorio, E. Strambini, G. Haack, M. Campisi, F. Giazotto,
{\it Photonic Heat Rectification in a System of Coupled
Qubits}, Phys. Rev. Applied \textbf{15}, 054050 (2021).

\bibitem{zhang2021}  Y. C. Zhang, S. H. Su, {\it Thermal rectification and negative 
differential thermal conductance based on a parallel coupled
double quantum-dot}, Physica A \textbf{584}, 126347
(2021).

\bibitem{tesser2022} Ludovico Tesser, Bibek Bhandari, Paolo Andrea Erdman,
Elisabetta Paladino, Rosario Fazio, Fabio Taddei,
{\it Heat rectification through single and coupled quantum
dots}, New J. Phys. \textbf{24}, 035001 (2022).

\bibitem{scheibner2008} Scheibner R, K\"onig M, Reuter D, A D Wieck, C Gould,
H Buhmann, L W Molenkamp, {\it  Quantum dot as thermal
rectifier}, New J Phys, \textbf{10}, 083016 (2008).

\bibitem{kastner1992}
M. A. Kastner, {\it The single-electron transistor},
Rev. Mod. Phys. \textbf{64}, 849 (1992).

\bibitem{benenti2017} 
G. Benenti, G. Casati, K. Saito, and R. S. Whitney,
{\it Fundamental aspects of steady-state conversion of heat to work at the nanoscale},
Phys. Rep. \textbf{694}, 1-124 (2017).

\bibitem{loss1998} 
D. Loss and D. P. DiVincenzo, 
{\it Quantum computation with quantum dots},
Phys. Rev. A \textbf{57}, 120 (1998).

\bibitem{burkard2000}
G. Burkard, H. A. Engel, and D. Loss, 
{\it Spintronics and quantum dots for quantum computing and quantum communication},
Fortschr. Phys.  \textbf{48}, 965 (2000).

\bibitem{ng1988}
Tai Kai Ng and Patrick A. Lee, {\it On-Site Coulomb Repulsion and Resonant Tunneling},
Phys. Rev. Lett. \textbf{61}, 1768 (1988).

\bibitem{glazman1988}
L. I. Glazman, M.E. Raikh,  {\it Resonant Kondo transparency of a barrier with quasilocal impurity states}, Pisma Zh. Eksp. Teor. Fiz. \textbf{47} 378 (1988)  [1988 JETP Lett. \textbf{47} 452 (Engl. Transl.)]

\bibitem{goldhabergordon1998} D. Goldhaber-Gordon, H. Shtrikman, D. Mahalu, D.
Abush-Maggder, U. Meirav, M. A. Kastner, Nature {\bf 391}, 156 (1998).

\bibitem{cronenwett1998} S. M. Cronenwett, T. H. Oosterkamp, L. P.
Kouwenhoven, Science {\bf 281}, 540 (1998).


\bibitem{buttiker1986}
M. Buttiker,
{\it Four-Terminal Phase-Coherent Conductance}, Phys. Rev. Lett. \textbf{57}, 1761 (1986).

\bibitem{bulgakov1999} E. N. Bulgakov, K. N. Pichugin, A. F. Sadreev, P. Streda,
P. Seba, {\it Hall-Like Effect Induced by Spin-Orbit Interaction},
Phys. Rev. Lett. \textbf{83}, 376 (1999).

\bibitem{pichugin2000}  K.N. Pichugin, P. Streda, P. Seba, A.F. Sadreev, {\it Resonance behaviour of the Hall-like effect induced by spinorbit
interaction in a four-terminal junction}, Physica E
\textbf{6}, 727 (2000).

\bibitem{wei2022}  Miaomiao Wei, Bin Wang, Yunjin Yu, Fuming Xu,
Jian Wang, {\it Nonlinear Hall effect induced by internal
Coulomb interaction and phase relaxation process in a
four-terminal system with time-reversal symmetry}, Phys.
Rev. B \textbf{105}, 115411 (2022).

\bibitem{sun2000}  Quin-feng Sun, Jian Wang, Tsung-han Lin, {\it Control of
the supercurrent in a mesoscopic four terminal Josephson
junction}, Phys. Rev. B \textbf{62}, 648 (2000).

\bibitem{song1998}
A. M. Song, A. Lorke, A. Kriele, and J. P. Kotthaus, W. Wegscheider, M. Bichler,
{\it Nonlinear Electron Transport in an Asymmetric Microjunction: A Ballistic Rectifier},
Phys. Rev. Lett. \textbf{80}, 3831 (1998).

\bibitem{ashcroft} N. A. Ashcroft, N. D. Mermin, {\it Solid State Physics}, 1976.

\bibitem{meir1992}
Y. Meir and N. S. Wingreen, 
{\it Landauer Formula for the current through an interacting electron region},
Phys. Rev. Lett. \textbf{68}, 2512 (1992).

\bibitem{meir1994}
N. S. Wingreen and Y. Meir, 
{\it Anderson model out of equilibrium:
Noncrossing-approximation approach to transport through a quantum dot},
Phys. Rev. B \textbf{49}, 11040 (1994).


\bibitem{haug-jauho1996} 
H. Haug and A.-P. Jauho, \textit{Quantum Kinetics in Transport and Optics of Semiconductors, 
Second, Substantially Revised Edition}, Springer, Berlin, 2008.




\bibitem{eckern2020} 
 U. Eckern and K. I. Wysoki\'nski,
{\it Two- and three-terminal far-from-equilibrium thermoelectric nanodevices in the Kondo regime}, 
New J. Phys. \textbf{22}, 013045 (2020).

\bibitem{eckern2021} 
 U. Eckern and K. I. Wysoki\'nski,
{\it Charge and heat transport through quantum dots with local and correlated-hopping interactions},  Phys. Rev. Research \textbf{3}, 043003 (2021).
%

\bibitem{lavagna2015}
M. Lavagna,
{\it Transport through an interacting quantum dot driven out-of-equilibrium}
J. Phys. Conf. Ser. \textbf{592}, 012141 (2015).


\bibitem{wysokinskimm2012}
M. M. Wysoki\'{n}ski, 
{\it Thermoelectric Effect in the Normal Conductor-Superconductor Junction: A BTK Approach},
Acta Phys. Pol. A \textbf{122}, 758 (2012).

\bibitem{wysokinskimm2013}
M. M. Wysoki\'{n}ski and J. Spa{\l}ek,
{\it  Seebeck effect in the graphene-superconductor junction},
J. Appl. Phys. \textbf{113}, 163905 (2013).

\bibitem{wysokinskimm2014}
M. M. Wysoki\'{n}ski,
{\it Temperature Dependence of the Zero-Bias Conductance in the Graphene NIS Junction},
Acta Phys. Pol. A \textbf{126}, A36 (2014).

\bibitem{polkovnikov2002}
A. Polkovnikov,
{\it Kondo effect in d-wave superconductors},
Phys. Rev. B \textbf{65}, 064503 (2002).


\end{thebibliography}
\end{document}